\documentclass{article}
\usepackage{graphicx}
\usepackage{amsmath}
\usepackage{amsthm}
\usepackage{amssymb}
\usepackage{bbm}

\newcommand{\ket}[1]{\mbox{$ | #1 \rangle $}} 
\newcommand{\bra}[1]{\mbox{$ \langle #1 | $}} 
\newcommand{\braket}[1]{\ensuremath{\langle #1\rangle}}

\newcommand{\trace}{\ensuremath{\operatorname{tr}}}
\newcommand{\Ad}{\ensuremath{\operatorname{Ad}}}
\newcommand{\ie}{\emph{i.e.}}
\newcommand{\eg}{\emph{e.g.}}

\newtheorem{proposition}{Proposition}[section]
\newtheorem{problem}{Problem}[section]
\newtheorem{theorem}{Theorem}[section]

\begin{document}

\title{Truncated $\mathfrak{su}(2)$ moment problem \\ for spin and
  polarization states}   

\author{Moroder Tobias$^{1,2}$, Keyl Michael$^3$, L\"utkenhaus
  Norbert$^{1,2}$}

\maketitle

\begin{enumerate}
\item QIT Group, Institute for Theoretical Physics I, and Max-Planck
  \\ Research Group, University Erlangen-Nuremberg, Germany 
\item Institute for Quantum Computing, University of Waterloo, Canada
\item Institute for Scientific Interchange Foundation, Torino, Italy.
\end{enumerate}


\begin{abstract}
We address the problem whether a given set of expectation values is
compatible with the first and second moments of the generic spin
operators of a system with total spin $j$. Those operators appear as
the Stokes operator in quantum optics, as well as the total angular
momentum operators in the atomic ensemble
literature. We link this problem to a particular extension problem for
bipartite qubit states; this problem is closely related to the
symmetric extension problem that has recently drawn much attention in
different contexts of the quantum information literature. We are able
to provide \emph{operational, approximate} solutions for every large
spin numbers, and in fact the solution becomes exact in the limiting
case of infinite spin numbers. Solutions for low spin numbers are
formulated in terms of a hyperplane characterization, similar to
entanglement witnesses, that can be efficiently solved with
semidefinite programming.  
\end{abstract}

\section{Introduction}

The ``black magic'' calculus of quantum mechanics allows us to make
predictions about expectation values of certain measurement
outcomes; however these expectation values are surprising in the
following ways: On one hand we know from Bell's inequality
\cite{bell_inequality} that not all possible quantum mechanical
expectation values are compatible with a local hidden variable model;
on the other hand, not all expectation values which originate from a
non-signaling constrained probability theory are quantum
mechanical \cite{nonlocal_boxes}. Given the quantum mechanical
description of the measurement device, the question arises which
expectation values are compatible with quantum theory at all. The
quantum mechanical moment problem \cite{moyal, narcowich}, as well as
its truncated version \cite{narcowich} are famous paradigms of this sort of
question. An operational characterization of physical expectation
values is of course desirable, however, only few examples are known so
far. A well-known example is provided in the context of Gaussian states for 
systems with finite numbers of degrees of freedom. The first two
moments of the corresponding position and momentum operators are
compatible with quantum mechanics if and only if the Schr\"odinger-Robertson
uncertainty principle is fulfilled \cite{holevo}. The condition can be
written in terms of the covariance matrix \cite{covariance_matrix},
which allows---together with a vector formed by the mean values---an
operational low-dimensional description of all the quantum states
compatible with the given moments. In most of the literature about
Gaussian states this low-dimensional description is exploited, which
pinpoints the importance of such a result. By contrast satisfying the
Schr\"odinger-Robertson uncertainty principle is not sufficient to
guarantee the existence of a quantum state in the case of higher
moments \cite{collins}, and a straightforward extension of Gaussian
states might be non-trivial \cite{marcincewicz}.

Clearly, given a set of linearly dependent operators the corresponding
expectation values have to reflect the same linear dependence. In
addition, each expectation value has to be in the convex hull of the
spectrum of the corresponding operator. In order to exploit further structures
of the set of operators, we consider operator sets with an underlying
Lie algebra structure. In the following we consider operators which are
irreducible representations of the Lie algebra $\mathfrak{su}(2)$, the
three-dimensional vector space of all tracefree, $2 \times 2$ complex matrices
$X$ with $X^\dag=-X$. The corresponding irreducible representation on the
Hilbert space $\mathcal{H}_j$ of dimension $d=2j+1$ is denoted by
$\partial \pi_j: \mathfrak{su}(2) \to \mathcal{B}(\mathcal{H}_j)$. In
addition, we use the label $\mathcal{L}^{(j)}_k=\partial
\pi_j(\sigma_k/2)$ for an irreducible representation of the Pauli operators
$\sigma_k$, which constitute a basis for the Lie algebra
$\mathfrak{su}(2)$. This set of operators satisfies the commutation
relations  
\begin{equation}
  [ \mathcal{L}^{(j)}_k, \mathcal{L}^{(j)}_l ] = i \varepsilon_{klm}
  \mathcal{L}^{(j)}_m
\end{equation}
where $\varepsilon_{klm}$ denotes the Levi-Civita tensor. These
commutation relations are well-known to be satisfied for the spin
operators, and indeed the set of operators $ \mathcal{L}^{(j)}_k$ can
be considered as the familiar \emph{spin operators} of a total
spin $j$, and we use the term in the following.  In fact, the
considered scenario appears in a variety of 
different fields in physics: The macroscopic spin measurements of a state of
an ensemble of $N$ two-level atoms only supported on the symmetric subspace
are spin operators of a spin $j=N/2$ system \cite{korbicz}; Any 
Stokes operator acting on a two mode system with fixed total photon number $N$
is described by a similar formalism \cite{korolkova}, however in contrast to
the spin operators the Stokes operators differ by a factor of $2$ in the
definition of the commutation relation, hence $\mathcal{\tilde
L}^{(N/2)}_k=2\mathcal{L}^{(N/2)}_k$. Note, since there is only one
irreducible representation of the Lie algebra $\mathfrak{su}(2)$ in a
given, fixed dimension $d$, the spin operators are
unique up to unitary transformations. In addition, the spin operators
satisfy the Casimir identity given by
\begin{equation}
  \label{eq: casimir}
  (\mathcal{L}_1^{(j)})^2 + (\mathcal{L}_2^{(j)})^2 +
  (\mathcal{L}_3^{(j)})^2 = j(j+1) \mathbbm{1}.  
\end{equation}

In the following, we are interested in the expectation values of
products of \emph{two} spin operators only. Although higher moments can be
measured in principle, in experiments it is often very tedious to get
accurate values for those moments, \eg, Ref.~\cite{lorenz}. Let us
formally denote the set of density operators $\rho$ on a given Hilbert space
$\mathcal{H}$ by 
\begin{equation}
  \mathcal{D}(\mathcal{H})= \left\{ \rho \in \mathcal{B}(\mathcal{H})
    \Big| \rho \geq 0, \trace(\rho)=1 \right\}.
\end{equation}
The problem stated below asks for the compatibility of a given set of
expectation values with spin operator measurements on a system of
fixed dimension\footnote{The considered set of given expectation values
  stands for knowing the first and second moments of the spin operators in
  arbitrary directions of the coordinate frame.}.

\begin{problem}[Truncated $\mathfrak{su}(2)$ moment
  problem]\label{problem1} Consider the set of operators
  $\mathcal{L}^{(j)}_k \mathcal{L}^{(j)}_l$ with 
  $k,l \in \{1,2,3\}$ formed by the products of \emph{two} spin operators
  acting on the Hilbert space $\mathcal{H}_j$ with dimension 
  $d=2j+1$. Given a set of expectation values $M_{kl} \in 
  \mathbbm{C}$ with $k,l \in \{1,2,3\}$, under which conditions do
  these expectation values originate from a quantum mechanical state, \ie,
  $\exists\: \rho \in \mathcal{D}(\mathcal{H}_j)$ such that
  $\trace(\mathcal{L}^{(j)}_k \mathcal{L}^{(j)}_l \rho)=M_{kl}, \forall k,l$? 
\end{problem}

We like to provide an \emph{operational} description of the set of valid
expectation values that enables working with the moments directly rather than
using the complete density matrix. In order to check for compatibility of a
given set of expectation values $M_{kl}$, the following solution is provided:
First, one verifies the linear dependence imposed by the Casimir
identity. Next, one reconstructs a particular operator $\rho_j(M)$ which acts
on the symmetric subspace of two qubits. 
The given expectation values are consistent with the spin operators if
and only if the operator $\rho_j(M)$ represents a valid density
operator for two qubits, that has exactly $2j-2$ Bose-symmetric
extensions. Although this reformulation does not provide an
operational description yet, it opens the possibility to apply results
from this extension problem directly to Problem
\ref{problem1}. Consequently, one can formulate \emph{operational
  approximations} to the truncated moment problem. Whenever one finds
a separable two qubit state $\rho_j(M)$, then it can be assured that
the corresponding expectation values are quantum mechanical, while if
one detects non-positivity of a certain operator $\tau_j(\rho_j(M))
\not \geq 0$, that depends on the total spin number $j$ and the
reconstructed two-qubit state $\rho_j(M)$, then the expectation values
are incompatible with quantum mechanics. This approximate
characterization gets more accurate the larger the total spin number
$j$ becomes, and converges to the exact solution in the case of
infinite spin numbers.  

The concept \emph{expectation value matrix} is introduced in
Sec. \ref{section2}, which imposes already a strong condition on
quantum mechanical expectation values in general. The Lie group structure
simplifies the problem to a particular standard form of the given expectation
value matrix. In Sec. \ref{section3}, we relate Problem \ref{problem1} to
the characterization of Bose-symmetric extendible two
qubit states by using a particular representation of the spin
operators. In addition, we show how this idea directly provides an
operational solution to a simplified version of the truncated moment
problem. Since the exact solution to this extension problem 
is unknown, two different methods are considered in order to provide solutions
for large (Sec. \ref{section4}) and small spin numbers
(Sec. \ref{section5}). In particular, Sec. \ref{section4} deals with the two
different approximation methods which both converge in the limit of high spin
numbers. In Sec. \ref{section5}, a solution to problems of the kind
like Problem \ref{problem1} in terms of hyperplanes is provided, 
which can be efficiently solved for low spin numbers $j$. A graphical 
comparison between the different sets is given in Sec. \ref{section6} that
demonstrates already the convergence. In Sec. \ref{section7} we summarize and
give an outlook on possible further directions.

\section{Expectation value matrix}
\label{section2}

Each expectation value $M_{kl}$ in Problem \ref{problem1} is labeled
by two indices, therefore each expectation value constitutes an
entry of a particular matrix, which we term \emph{expectation value matrix} 
in the following. Considering the set of expectation values
$M$ in matrix form already enables us to derive a strong statement
about quantum mechanical expectation values in general; it even holds
without the Lie group structure and has already appeared in the
literature \cite{vogel, miranowicz, korbicz1, rigas, haeseler}. The
construction of the expectation value matrix is equivalent to the
derivation of the Schr\"odinger-Robertson uncertainty principle, 
which all given expectation values must clearly satisfy in order to
originate from a quantum state.

\subsection{Definition \& Properties}

The expectation value matrix defines a linear map on the set of density
operators, which preserves hermiticity and positive
semidefiniteness. Therefore every expectation value matrix $M$, which
originates from a valid quantum state $\rho$, \ie, $M=M(\rho)$, must
necessarily be positive semidefinite. The following proposition
introduces a slightly bigger \emph{expectation value
  matrix}\footnote{We use the term \emph{expectation value matrix} for
both, the $3 \times 3$ matrix $M$ and the $4 \times 4$ matrix
$\chi$. It should be clear form the context to which matrix one refers.} $\chi:
\mathcal{B}(\mathcal{H}_j) \to \mathcal{B}(\mathbbm{C}^4)$, which
contains our original expectation value matrix $M$ in form of a submatrix,
hence the given statements follows automatically. We skip the proof
since it has already appeared in the literature, \eg,
Ref. \cite{miranowicz}. Note that further constraints are naturally imposed if
  the corresponding operator set shows linear dependence.
Let us mention that Gaussian states are examples where the
conditions on the expectation value matrix are necessary and sufficient for
the existence of a quantum state. 

\begin{proposition}[Expectation value matrix]\label{prop: evm} Let
  $\mathcal{F}$ denote the set of operators acting on the Hilbert
  space $\mathcal{H}_j$ formed by the identity operator and the spin
  operators. The \emph{expectation value matrix} $\chi: 
  \mathcal{B}(\mathcal{H}_j) \to \mathcal{B}(\mathbbm{C}^4)$ is defined by
  \begin{equation}
    \mathcal{B}(\mathcal{H}_j) \ni A \mapsto \chi_{kl}(A)=\trace(F_k^\dag F_lA)
  \end{equation}
  with $F_k \in \mathcal{F}$ for $k=1,\dots, 4$. This expectation value
  matrix $\chi$ has the following properties:
  \begin{itemize}
  \item $\chi(aA+bB)=a \chi(A) + b \chi(B)$ for $A,B \in
    \mathcal{B}(\mathcal{H}_j),\;a,b \in \mathbbm{C}$.
  \item If $A=A^\dag$ then $\chi(A)=\chi(A)^\dag$. 
  \item If $A \geq 0$ then $\chi(A) \geq 0$.
  \end{itemize}
\end{proposition}

\subsection{Standard form}

In order to reduce the number of parameters in the expectation
value matrix $M$ we give a \emph{standard form} first. The standard form
allows us to consider expectation value matrices with purely imaginary
off-diagonal entries only, hence the set of expectation values is solely
characterized by the mean values and the variances of the spin operators
$\mathcal{L}^{(j)}_k$. In order to derive the standard form one has to
show the covariance property of the expectation value matrix under the
adjoint representation of the corresponding Lie group
$\operatorname{SU(2)}$, given by $\Ad: \operatorname{SU(2)} \to
\mathcal{B}({\mathfrak{su}(2)})$ and $\Ad_U(x)= U^\dag x U$, for $U
\in \operatorname{SU(2)}$ and $x \in \mathfrak{su}(2)$. Acting on the
$\mathbbm{C}^3$, this adjoint representation is unitary equivalent to
the irreducible spin-1 representation $\pi_1$ of the Lie group
$\operatorname{SU(2)}$. 
This allows us to diagonalize the real part of the expectation value matrix. 
The idea is quite common in the literature of atomic ensembles,
see, \eg, Ref.~\cite{toth}, hence the proof is omitted.  

\begin{proposition}[Covariance property \& Standard form] The
  expectation value matrix $M=M(\rho)$ satisfies the covariance property 
  \begin{equation}
    M(\pi_j(U)\:\!\rho\:\! \pi_j(U)^\dag)=\pi_1(U)\:\! M(\rho)\:\!
    \pi_1(U)^\dag,\;\; \forall U \in \operatorname{SU(2)}
  \end{equation}
  where $\pi_j$ denotes the irreducible representation of the group
  $\operatorname{SU(2)}$. Therefore each expectation value matrix $M$ can be
  transformed to the \emph{standard form}
  \begin{equation}
    \label{eq: standardform}
    M=D+iA,
  \end{equation}
  where $D$ is a diagonal matrix with fixed trace,
  $\trace(D)=j(j+1)$ and $iA$ denotes the antihermitean, tracefree part. 
\end{proposition}

\section{Reduction}
\label{section3}

Each irreducible representation $\partial \pi_j$ of the Lie algebra
$\mathfrak{su}(2)$ can be built up from its two-dimensional,
fundamental representations $\partial \pi_{1/2}$ by acting on a
restricted subspace of the tensor product space of multiple qubits
\cite{hall}. Such a particular form of the representation appears as the
usual representation in the case of atomic ensembles
\cite{korbicz}. Given a valid density operator on this restricted
tensor product space of multiple qubits, only an effective two qubit
state suffices to determine the expectation values of the products of
at most \emph{two} Lie algebra elements,  cf. Ref. \cite{wang}. Using
the prescribed formalism enables us to establish a connection between Problem
\ref{problem1} and the Bose-symmetric extendibility question for a
particular two qubit state, which constitutes the first result. This
particular extension problem shows similarities to the so-called
symmetric extension problem, which recently has drawn attention in the
literature \cite{doherty, doherty1, terhal, morodersym,
  wolf}. Moreover, we further demonstrate the reduction idea by
solving a simplified version of the truncated moment problem.

\subsection{Group theory}

This part reviews the irreducible representation $\partial \pi_j$ of
the Lie algebra $\mathfrak{su}(2)$ on the tensor product
space of multiple qubits, restricted to the symmetric
subspace. The symmetric subspace
$\mathcal{H}_+^{\otimes 2j}$, indicated by the
  subscript $+$, consists of all vectors $\ket{\psi} \in
\mathcal{H}^{\otimes 2j}$, with $\mathcal{H}=\mathbbm{C}^2$, that are
invariant under all possible permutations $\pi_{2j} (p)
\ket{\psi}=\ket{\psi}, p \in \operatorname{S}_{2j}$. Here
$\pi_{2j}$ denotes an irreducible representation of
the permutation group $\operatorname{S}_{2j}$, which
acts on a basis of product states as $\pi_{2j}(p) \ket{i_1}\otimes \dots \otimes
  \ket{i_{2j}}=\ket{i_{p^{-1}(1)}}\otimes \dots \otimes
  \ket{i_{p^{-1}(2j)}}$. The following proposition describes the 
particular form of the representation.

\begin{proposition}[Irreducible $\mathfrak{su}(2)$ representation] One has:
  \begin{itemize}
  \item  The two Hilbert spaces $\mathbbm{C}^{2j+1}$ and
  $\mathcal{H}_+^{\otimes 2j}$ with $\mathcal{H}=\mathbbm{C}^2$ are
  isomorph.
  \item The irreducible representation of the Lie algebra $\partial \pi_j:
  \mathfrak{su}(2) \to \mathcal{B}(\mathcal{H}_+^{\otimes 2j})$ is given by
  \begin{equation}
    \partial_j \pi (x) = \sum_{k=1}^{2j} x^{(k)}
    \bigg |_{\mathcal{H}_+^{\otimes 2j}}, \;\; x \in \mathfrak{su}(2),
  \end{equation}
  with $x^{(k)}=\mathbbm{1}^{\otimes k-1} \otimes \partial
  \pi_{1/2}(x) \otimes \mathbbm{1}^{\otimes 2j-k}$, $\partial
  \pi_{1/2}(x)$ is the two dimensional fundamental
  representation, and $|_{\mathcal{H}_+^{\otimes 2j}}$ denotes the
  restriction to $\mathcal{H}_+^{\otimes 2j}$. 
  \end{itemize}
\end{proposition}

\subsection{Two qubit reduction}

Since the two abstract Hilbert spaces of the previous section are isomorph, any
density operator of a spin $j$ system can be interpreted as the density
operator of a particular multipartite system of $2j$ qubits, \ie,  $\rho_j \in
\mathcal{D}(\mathcal{H}_+^{\otimes 2j})$. In the following we are only
interested in the expectation values of products of at most two spin
operators. Because of the particular symmetry of the multipartite qubit
state and the explicit representation of these operators, these
expectation values depend only on an effective spin-$1$ system. 
The following proposition summarizes this reduction and links the
corresponding spin operators. This reduction idea
already appears in the context of atomic ensembles; a proof
can be found, \eg, in Ref. \cite{wang}, which contains also examples
of these reduced states.  

\begin{proposition}[Spin-$1$ reduction] \label{prop: spin1state}
  For every density operator $\omega \in
  \mathcal{D}(\mathcal{H}^{\otimes 2j}_+)$, the expectation values of
  the products of at most two Lie algebra elements only need to
  be calculated on an effective spin-1 state $\rho_j=\trace_{3\dots
  2j}(\omega) \in \mathcal{D}(\mathcal{H}^{\otimes 2}_+)$ and are
  given by $\trace(\partial \pi_j(x) \omega)=\trace(\Lambda(\partial
  \pi_j(x))\rho_j)$ and similar $\trace (\partial_j \pi (x)^\dag
  \partial \pi_j (y) \omega) = \trace (\Lambda(\partial_j \pi
  (x)^\dag \partial \pi_j(y)) \rho_j)$, for $x,y \in
  \mathfrak{su}(2)$ with the operators given by \footnote{Note,
      $\Lambda(\cdot)$ is used as a symbol here and should not be
      interpreted as a map $\Lambda:
      \mathcal{B}(\mathcal{H}_+^{\otimes 2j}) \to 
      \mathcal{B}(\mathcal{H}_+^{\otimes 2})$.}   
  \begin{equation}
    \label{eq: operator_one} 
    \Lambda(\partial \pi_j(x)) = \left( 2j \;\partial \pi_{1/2}(x)
    \otimes \mathbbm{1} \right) \bigg|_{\mathcal{H}_+^{\otimes 2}},
  \end{equation}
  for the mean values, and similar for the products of two spin
  operators 
  \begin{eqnarray}
    \label{eq: operator_two}
    \Lambda(\partial \pi_j (x)^\dag \partial \pi_j (y)) &=& \left( 2j\;
    \partial \pi_{1/2} (x)^\dag \partial \pi_{1/2}
    (y) \otimes \mathbbm{1} \right. \\ &&     
    \nonumber 
    \left. + 2j (2j-1)\; \partial \pi_{1/2} (x)^\dag \otimes
    \partial \pi_{1/2} (y)) \right)
  \bigg|_{\mathcal{H}_+^{\otimes 2}}.  
  \end{eqnarray}
\end{proposition}
 
Only the reduced spin-$1$ state $\rho_j=\trace_{3\dots 2j}(\omega)$
of a given multipartite \mbox{state $\omega$} of total spin $j$ determines the
expectation values of products of two spin operators. However not all
possible spin-$1$ density operators are reduced states of such
multipartite states. The formal definition of valid two
qubit reductions of a spin $j$ system is given by  
\begin{equation}
  \label{eq:exact}
  \mathcal{S}_j= \left\{\rho \in \mathcal{D}(\mathcal{H}_+^{\otimes 2})
    \Big | \exists \omega \in \mathcal{D}(\mathcal{H}_+^{\otimes 2j}),
    \text{tr}_{3\dots 2j} (\omega) = \rho \right\},
\end{equation}
with $j\geq 1$. Hence, a spin-$1$ state can correspond to a spin $j$ system if
and only if it can be extended to a multipartite state of $2j$ qubits
which has support only on the symmetric subspace. This definition
resembles the notion of \emph{Bose-symmetric
extensions} for bipartite states, which is closely related to the
symmetric extensions for bipartite states, cf. Ref \cite{doherty}
\footnote{Note, that in the definition of symmetric
    extensions, as given in Ref. \cite{doherty}, the corresponding
    multipartite state must be invariant under the permutation of
    the individual subsystems, while in the case of
    \emph{Bose-symmetric} extensions one requires for the multipartite
    state to have support on the symmetric subspace only.
    Although both problems are very similar---and in fact,
    many results can be ``borrowed'' directly  form the symmetric
    extension case---certain properties differ. For example, every
    separable, permutation invariant density operator has a symmetric
    extension to arbitrary many copies, while it does not need to have
    a Bose-symmetric extension.}.   
The next theorem establishes the connection between the Problem
\ref{problem1} and the problem of Bose-symmetric
extensions of a particular spin-$1$ state.   

\begin{theorem}[Bose-symmetric Extensions $\Leftrightarrow$  Quantum
  mechanical expectation values] \label{thm: extension} Given the
  expectation value matrix $M$ the corresponding spin-$1$ state
  $\rho_j=\rho_j(M)$ is uniquely determined by  
  \begin{equation}
    \trace(\Lambda(\mathcal{L}^{(j)}_k\mathcal{L}^{(j)}_l ) \rho_j) = M_{kl},
  \end{equation}
  for all $k,l \in \{1,2,3\}$, and $\Lambda(\cdot)$ is given by Eq. \ref{eq:
  operator_two}. The given expectation value matrix $M$ is 
  \emph{quantum mechanical} if and only if $\rho_j(M) \in
  \mathcal{S}_j$, \ie, the state $\rho_j(M)$ has exactly $2j-2$
  Bose-symmetric extensions. 
\end{theorem}

\begin{proof}
It is straightforward to check that the operators given by
Eq. \ref{eq: operator_two} span an operator basis for a generic
spin-$1$ state, hence the corresponding spin-$1$ state
$\rho_j=\rho_j(M)$ is completely determined by the expectation value
matrix $M$. This particular two-qubit state can correspond to a
spin $j$ system if and only if it is an element of the class
$\mathcal{S}_j$.  
\end{proof}

\subsection{First moment problem}

In the present subsection, we briefly discuss a simplified version of the
truncated moment problem in which one asks for compatibility of the first
moments only. This problem allows an operational solution and highlights again
the importance of the reduction idea. It states as follows:

\begin{problem}[First moment problem]
  Consider the set of spin operators $\mathcal{L}^{(j)}_k$ with $k \in
  \{1,2,3\}$ acting on the Hilbert space $\mathcal{H}_j$ of dimension 
  $d=2j+1$. Given a set of expectation values $L_{k} \in 
  \mathbbm{R}$ with $k \in \{1,2,3\}$, under which conditions do
  these expectation values originate from a quantum mechanical state, \ie,
  $\exists\: \rho \in \mathcal{D}(\mathcal{H}_j)$ such that
  $\trace(\mathcal{L}^{(j)}_k \rho)=L_{k}, \forall k$? 
\end{problem}

The operational description of the set of valid expectation values relies on
the reduction to a spin-$1/2$ problem, in analogy to Proposition
\ref{prop: spin1state}, and that every valid spin-$1/2$ state represents a
possible reduced state of a particular spin-$j$ state, \ie, each single qubit
density operator can be extended to a Bose-symmetric density operator
of $2j$ qubits. The next proposition contains the operational solution.   

\begin{proposition}[Operational description] \label{prop:
    firstmoment}The expectation values 
  $L_k \in \mathbbm{R}$ with $k\in\{1,2,3\}$ are \emph{compatible} with the
  spin operators of a generic spin $j$ system if and only if it holds
  \begin{equation}
    \label{eq: condfirst}
    \sum_{i=1}^3 L_i^2 \leq j^2.
  \end{equation}
\end{proposition}

\begin{proof}
In analogy to Proposition \ref{prop: spin1state}, it holds that for every
spin-$j$ density operator $\omega \in \mathcal{D}(\mathcal{H}^{\otimes 2j}_+)$,
the expectation values of each Lie algebra element $\partial \pi_j(x)$ need
only to be calculated on an effective \emph{qubit} state $\tilde
\rho_j=\trace_{2\dots 2j}(\omega )$ and are given by $\trace(\partial
\pi_j(x) \omega)=\trace(\tilde \Lambda(\partial \pi_j(x)) \tilde
\rho_j)$ with $\tilde \Lambda(x)= 2j \partial \pi_{1/2}(x)$. Contrary
to the spin-$1$ case, every qubit state $\tilde \rho$ can be extended
back to a multipartite state $\omega=\tilde \rho^{\otimes 2j}
|_{\mathcal{H}_+^{\otimes 2j}}$, \ie, to a density operator of a spin $j$ 
system with the correct first moments. Thus, the given expectation values
$L_k$ only have to form a valid qubit density operator $\tilde \rho_j(\{L_k\})
\in \mathcal{D}(\mathcal{H})$. The set of given expectation values $L_k$
uniquely determines the Bloch vector $L_k/j$ of the reduced qubit
state $\tilde \rho_j$, which is quantum mechanical if and only if Eq. \ref{eq:
condfirst} holds.  
\end{proof}

We close with a discussion about possible solution states to the first
moment problem. Given the set of expectation values, any compatible quantum
state can be parameterized as $\rho({\bf{x}})=\rho_{\text{fix}}(\{L_k\}) +
\rho_{\text{open}}({\bf{x}})$, with a fixed part
$\rho_{\text{fix}}(\{ L_k\})=\mathbbm{1}/(2j+1) +
  \sum_k a_k \mathcal{L}^{(j)}_k$ and $a_k=L_k/ \| \mathcal{L}^{(j)}_k
  \|_2^2$, that is completely determined by the first moments $L_k$, and an
orthogonal open part $\rho_{\text{open}}({\bf{x}})$, which linearly depends on
some open parameters ${\bf{x}}$, $x_i \in \mathbbm{R}$, and which have to be
chosen such that $\rho({\bf{x}}) \geq 0$ forms a valid density operator. 
Proposition \ref{prop: firstmoment} guarantees existence of such a set of open
parameters ${\bf{x}}$ as long as the given expectation values fulfill
Eq.~\ref{eq: condfirst}. A special class of solutions
constitutes the case in which $\rho_{\text{fix}}(\{ L_k \}) \geq 0$
forms already a positive semidefinite operator itself. However, this
is the case if and only if $\sum_{i=1}^3 L_i^2 \leq (j+1)^2/9$, which
shows that the open part $\rho_{\text{open}}({\bf{x}})$ is indeed
necessary since it allows compensation of a non-positive fixed part
$\rho_{\text{fix}}(\{ L_k \}) \not \geq 0$ in certain cases.

\section{Approximating sub- and superset}
\label{section4}

Any solution to the Bose-symmetric extension problem
for two qubits constitutes a solution for the $\mathfrak{su}(2)$
moment problem; however, an analytic characterization of those sets appears
cumbersome. 
Therefore, two \emph{operational approximation} methods to
the set of extendible states are described in the following, which
allows identification of certain expectation value matrices, that are
either quantum mechanical (subset) or not (superset). Both
approximating sets converge to the exact set in the 
limit of infinite numbers of extensions. The subset characterization
relies on a result by \textit{Doherty et al.}, Ref.~\cite{doherty}, 
while the supersets depend on the expectation value matrix. 
Both approximation sets (and the corresponding convergence statements) are
given in terms of renormalized expectation values of an
expectation value matrix defined by 
\begin{equation}
  \label{eq: renormalized}
  u_k\equiv\frac{\braket{\mathcal{L}^{(j)}_k}}{j}, \;\;
  v_k\equiv\frac{\braket{(\mathcal{L}^{(j)}_k)^2}}{j(j-1/2)}-\frac{1}{2j-1},
\end{equation}
for $k\in\{1,2,3\}$. Note that one has only to consider these $6$
expectation values because of the standard form of the expectation
value matrix as described by Eq. \ref{eq: standardform}. More
precisely, the parameters $v_k$ determine the diagonal entries of the
expectation value, while $u_k$ fixes the antihermitean, tracefree part.
These renormalized expectation values have the advantage that the determined
spin-$1$ state $\rho_j(\{u_k, v_k\})$ is independent of the considered
total spin number $j$.   

\subsection{Approximating subset}

The set of separable two qubit states, supported on the symmetric
subspace, forms an approximating subset to the set of Bose-symmetric
extendible states, independent of the number of 
extensions. Hence, if a given expectation value matrix $M$ allows the
reconstruction of a separable spin-$1$ state $\rho_j(M)$, then it can
be assured that all expectation values in $M$ are quantum
mechanical. The set of separable spin-$1$ states is given by 
\begin{equation}
  \label{eq:inner}
  \mathcal{R}=\left\{ \rho \in \mathcal{D}(\mathcal{H}_+^{\otimes
      2}) \Big | \rho^\Gamma \geq 0 \right\},
\end{equation}
where $\Gamma$ denotes the partial transpose, which is necessary and
sufficient to characterize separable states in these low
dimensions \cite{peres},\cite{entanglement_witness}. The next theorem outlines
the properties of this subset. The convergence result is
a direct consequence of a corresponding result for states 
which have a symmetric extension \cite{doherty}. For a fixed number of
photons, any convex combination of polarization states, where all
photons are in one optical mode, constitutes physical relevant examples
where the qubit density operator is separable. In contrast, any
spin squeezed state must have an entangled two qubit density operator
\cite{wangsanders}, as well as certain multiqubit states in the atomic
ensemble literature \cite{stokten}.

\begin{theorem}[Approximating subset] The sets $\mathcal{S}_j$ with $j
  \geq 1$, and $\mathcal{R}$ given by Eqs. \ref{eq:exact} and
  \ref{eq:inner} respectively satisfy:
  \begin{enumerate}
  \item If $\rho \in \mathcal{R}$, then $\rho \in \mathcal{S}_j,
    \forall j$. Hence $\mathcal{R} \subseteq \mathcal{S}_j, \forall j$.
  \item If $\rho \in \mathcal{S}_j$, then $\rho \in
      \mathcal{S}_{j^\prime}$ with $j^\prime \leq j$. Hence
      $\mathcal{S}_j \subseteq \mathcal{S}_{j^\prime}$ for $j^\prime
      \leq j$. 
  \item $\lim_{j \to \infty} \mathcal{S}_j \equiv
      \bigcap\limits_j \mathcal{S}_j \subseteq \mathcal{R}$. 
  \end{enumerate}
\end{theorem}

\begin{proof}
  Any separable density operator $\rho \in \mathcal{R}$ can be
  written as a convex combination of pure product states. We assume
  the following decomposition $\rho=\sum_i p_i
  \ket{\alpha_i}\bra{\alpha_i} \otimes \ket{\beta_i}\bra{\beta_i}$,
  where $\{ p_i \}$ denotes the probability distribution
  and $\ket{\alpha_i}, \ket{\beta_i} \in \mathcal{H}, \forall
  i$. Because of $\rho \in \mathcal{B}(\mathcal{H}_+^{\otimes 2})$, it
  must hold that $\trace(P_-\rho)=0$, where $P_- \geq 0$ denotes the
  projector onto the antisymmetric subspace. Given the particular
  decomposition into positive semidefinite operators it follows that
  the trace must vanish for each term, 
  $\trace(P_- \ket{\alpha_i}\bra{\alpha_i} \otimes
  \ket{\beta_i}\bra{\beta_i})=0, \forall i$, which shows that 
  each $\ket{\alpha_i}\otimes \ket{\beta_i} \in \mathcal{H}_+^{\otimes
    2}$. This can be the case iff $\ket{\alpha_i}=\ket{\beta_i},
  \forall i$. This allows us to write down a valid Bose-symmetric
  extension given by $\omega=\sum
  p_i\ket{\alpha_i}\bra{\alpha_i}^{\otimes 2j}$, which then proves the
  first statement, $\rho \in \mathcal{S}_j$. 
  
  The second statement follows trivially. The last
  point is a direct consequence of Theorem $1$ in Ref. \cite{doherty},
  which states that any bipartite mixed states $\rho_{AB} \in
  \mathcal{D}(\mathcal{H}_A \otimes \mathcal{H}_B)$, which has
  arbitrary many symmetric extensions to one of the subsystems must
  necessarily be separable. 
\end{proof}

\subsection{Approximating superset}

Exploiting the characteristics of the expectation value matrix allows us to
formulate an approximating superset to the set of extendible
states. Any reduced spin-$1$ state $\rho_j=\trace_{3\dots2j}(\omega)$ of a
given state $\omega \in \mathcal{D}(\mathcal{H}_+^{\otimes 2j})$ allows the
computation of the expectation values of a corresponding spin $j$ system via
the operator sets $\Lambda(\cdot)$ given by Eqs. \ref{eq: operator_one},
\ref{eq:   operator_two}; in this sense the expectation values from a spin $j$
system can be ``recovered'' from the reduced two qubit state. Proposition
\ref{prop: rexp} introduces the \emph{reduced expectation value matrix}
$\tau_j$, which tries to reproduce a normalized version of
the original expectation value matrix $\chi$ of the spin $j$ state
only from the reduced spin-$1$ state; renormalization protects against
divergence of the expectation values in the limit $j \to
\infty$. It holds that if the two qubit density operator $\rho$
has a valid extension to $2j$ qubits, then one obtains a positive
semidefinite operator.  

\begin{proposition}[Reduced expectation value matrix] \label{prop: rexp} Let
  $\mathcal{F}$ 
  denote the set of operators acting on the Hilbert space
  $\mathcal{H}_j$ formed by the identity and the renormalized spin
  operators $\mathcal{L}^{(j)}_k / j$. The
  \emph{reduced expectation value matrix} of order $j$, denoted by $\tau_j:
  \mathcal{B}(\mathcal{H}_+^{\otimes 2}) \to
  \mathcal{B}(\mathbbm{C}^{4})$, is defined as
  \begin{equation}
    \mathcal{B}(\mathcal{H}^{\otimes 2}_+) \ni A \mapsto
    \tau_j(A)_{kl}= \trace(\Lambda(F_k^\dag F_l) A), 
  \end{equation}
  with $F_k \in \mathcal{F}$ for $k=1, \dots 4$, and the operators
  $\Lambda(\cdot)$ are given by Eqs. \ref{eq: operator_one}, \ref{eq: 
  operator_two}. The map is linear and preserves hermiticity. In
  addition, it holds that if $\rho \in \mathcal{S}_j$ then 
  \begin{equation}
    \tau_j(\rho) \geq 0.
  \end{equation}
\end{proposition}

The approximating superset consists of all spin-$1$
states for which the reduced expectation value matrix of order $j$
delivers a positive semidefinite operator, and is denoted by
\begin{equation}
  \mathcal{T}_j= \left\{\ \rho \in \mathcal{D}(\mathcal{H}_+^{\otimes
      2}) \Big | \tau_j(\rho) \geq 0 \right\},
\end{equation}
for $j \geq 1$. Unlike the subset characterization, any superset
depends on the number of considered extensions $j$. The next theorem
lists the properties of the approximating superset. If a given
reconstructed spin-$1$ operator $\rho_j(M)$ is  
not an element of the class $\mathcal{T}_j$, then the corresponding
expectation values in $M$ are incompatible with quantum mechanics.  
The supersets converge to the set of separable
states in the limit of infinite number of extensions, hence
the reduced expectation value matrix delivers an alternative
necessary and sufficient entanglement criterion, valid \emph{only} for
symmetric two qubit state, but which does not rely on partial
transposition. However, in order to prove convergence, we exploit a
necessary and sufficient criterion for entanglement of the reduced two
qubit state in terms of the expectation values of the spin operators,
cf. Ref.~\cite{korbicz2}, which is derived using the partial transposition.

\begin{theorem}[Approximating superset] \label{thm: outer} The sets
  $\mathcal{T}_j$ with $j\geq 1$ satisfy: 
  \begin{enumerate}
  \item If $\rho \in \mathcal{T}_j$, then $\rho \in \mathcal{T}_{j^\prime}$
    with $j^\prime \leq j$. Hence $\mathcal{T}_j \subseteq
    \mathcal{T}_{j^\prime}$ for $j^\prime \leq j$.
  \item If $\rho \in \mathcal{S}_j$, then $\rho \in \mathcal{T}_j,
    \forall j$. Hence $\mathcal{S}_j \subseteq \mathcal{T}_j, \forall j$.
   \item $\lim_{j \to \infty} \mathcal{T}_j \equiv
       \bigcap\limits_j \mathcal{T}_j \subseteq \mathcal{R}$. 
  \end{enumerate}
\end{theorem}

\begin{proof}
  The proof of the first two statements is straightforward. Convergence
  is proven via contradiction along the following line: First, one
  uses the result given in Ref.~\cite{korbicz2} to provide a necessary
  and sufficient criterion for two qubit entanglement in terms of the
  renormalized expectation values. Second, one derives a necessary 
  condition for non-negativity of the reduced expectation value map $\tau_j$
  in the limit $j\to\infty$, expressed again in terms of the renormalized
  expectation values of the spin-$1$ state. The two given conditions are
  mutually exclusive, hence no entangled state can be part of the
  superset $\mathcal{T}_j$ in the limit $j \to
  \infty$. This shows convergence to the separable states in the end,
  since separable spin-$1$ states are trivially part of the outer
  approximation. 
  
  For a given direction $\hat n \in \mathbbm{R}^3$ the corresponding
  spin operator in this direction is defined by $\mathcal{L}^{(j)}_{\hat
  n}=\partial \pi_j(\hat n \cdot \vec {\sigma}/2)$, and $\vec {\sigma}$
  denotes the vector of Pauli matrices. Let $u_{\hat n}$ and $v_{\hat n}$
  denote the corresponding renormalized expectation values of this spin
  operator given by Eq. \ref{eq: renormalized}. Using the result given in
  Ref. \cite{korbicz2}, the reduced two-qubit state $\rho_j$ only supported on
  the symmetric subspace is entangled iff there exists a direction $\hat n$,
  such that $v_{\hat n} - u_{\hat n}^2 < 0$ holds. 

  The reduced expectation value matrix of order $j$ of a given spin-$1$
  density operator $\rho$, $\tau_j(\rho)$ is positive semidefinite
  iff all principle minors are non-negative. Using the
  unitary freedom of the matrix representation, in combination with
  the non-negativity of all possible $2\times 2$ submatrices ensures that
  $\text{Var}(\mathcal{L}^{(j)}_{\hat n})/j^2 \geq 0$ for all possible
  directions $\hat n$. If one re-expresses this condition in the
  normalized expectation values one arrives at 
  \begin{equation}
    v_{\hat n} - u_{\hat n}^2 +\frac{1}{2j}(1-v_{\hat n}) \geq 0,
  \end{equation}
  which becomes $v_{\hat n} - u_{\hat n}^2 \geq 0$ in the limit $j\to
  \infty$. This condition ensures that the entanglement condition can
  never be met in the limit, and this proves the convergence.
\end{proof}

\section{Hyperplane characterization}
\label{section5}

In this part we discuss some general features of compatibility
problems, in which one asks whether a certain set of expectation
values are consistent with an operator set, similar to Problem
\ref{problem1}.  Consider a set of linear operators on a finite
dimensional Hilbert space. Without loss of generality, only a set of
linear independent, hermitian operators needs to be considered,
because any operator can be decomposed into hermitian operators, and
linear dependencies in the operator set demand only trivial conditions
on the corresponding expectation values.  

\begin{problem}[Consistency problem] \label{problem2} Consider a set
  of linearly independent, hermitian operators acting on a finite
  dimensional Hilbert space $\mathcal{H}$, $\{A_i\}$ with
  $i=1,\dots,n$. Given an \emph{expectation value vector} ${\bf{b}}
  \in \mathbbm{R}^n$, under which conditions do these expectation
  values originate from a quantum mechanical state, \ie,  $\exists\:
  \rho \in \mathcal{D}(\mathcal{H})$ such that $\trace(A_i \rho)=b_i,
  \forall i$?  
\end{problem}

Since the set of density operators $\mathcal{D}(\mathcal{H})$ is
compact, and because of linearity of the map $\mathcal{M}:
\mathcal{D}(\mathcal{H}) \to \mathbbm{R}^n$ that assigns the
corresponding expectation values to a given state, $\rho \mapsto
\mathcal{M}(\rho)_i=\trace(A_i \rho)$, the set of valid expectation
values is also compact.  According to the Hahn-Banach
theorem, every expectation value vector outside
this set can be separated by a corresponding hyperplane, in analogy to
entanglement witnesses \cite{entanglement_witness} for the 
separability problem. The following theorem summarizes the necessary
properties of those hyperplanes and therefore provides a possible
characterization of the set of quantum mechanical expectation
values. Given a particular expectation value vector, the search for
the optimal hyperplane can be cast in the form of a semidefinite
program \cite{vandenberghe}, which can be solved efficiently with
interior-point methods. The given proof is based on this idea and
employs well-known duality relations from semidefinite programming,
cf. Ref.~\cite{vandenberghe}. 

In the case of the considered $\mathfrak{su}(2)$ moment problem,
a formulation in terms of a semidefinite program offers an efficient
way to characterize the set of possible expectation values for low
total spin numbers $j$, \ie, in situations where the approximation
method fails to 
provide a good description. Note that if one uses the particular
representation of the considered spin operators, then one recovers the
characterization of Bose-extendible states in terms of witness
operators similar as described in Ref. \cite{doherty}.

\begin{theorem}[Hyperplane characterization]\label{thm: witness}
Given a set of linearly independent operators $\{ A_i \}$ and an
expectation value vector ${\bf{b}} \in \mathbbm{R}^n$, it holds: 
\begin{itemize}
\item The vector ${\bf{b}}$ is \emph{non-quantum mechanical}, if and only if
  there exists a hyperplane, characterized by the normal vector
  \mbox{$z \in \mathbbm{R}^n: Z=\sum z_i A_i\geq0$}, 
  $ \trace(Z)=1$, which \emph{detects} it $z^T{\bf{b}} < 0$.  
\item The vector ${\bf{b}}$ is \emph{quantum mechanical}, if and only
  if for all hyperplanes, characterized by the normal
    vector $z \in \mathbbm{R}^n: Z=\sum z_i A_i \geq 
  0,\trace(Z)=1$ it holds that $z^T{\bf{b}} \geq 0$. 
\end{itemize}
\end{theorem}

\begin{proof}
The Gram-Schmidt orthogonalization allows to transform any set of
arbitrary linear independent, hermitian operators $\{A_i\}$ to 
an orthonormal set of hermitian operators $\{ S^\prime_i \}$, \ie,
the operators satisfy $\trace(S_i S_j)=\delta_{ij}$ and
$\trace(S_i) \propto \delta_{i1}$. The connection is 1-to-1 and
the corresponding expectation value vector ${\bf{b}}$ has to be
transformed in the same way, which results in the new expectation
value vector ${\bf{t}}$.   

The set $\{S_i\}$ for $i=1,\dots n$ can be extended to a hermitian
operator basis $\{S_i\}$ with $i=1,\dots d^2$ and 
$d=\dim(\mathcal{H})$. 
This basis set $\{S_i\}$ satisfies again $\trace(S_i S_j)=\delta_{ij}$
and $\trace(S_i) \propto \delta_{i1}$, which states that the first
operator is proportional to the identity, $S_1=\mathbbm{1}/\sqrt{d}$,
and all other elements $S_j$ for $j>1$ are tracefree.  
Therefore every density operator can be
expressed as $\rho=\sum_i x_i S_i$ with $x_i=\trace(S_i \rho)$. It
must hold $x_i=t_i$ for $i=1,\dots n$ since otherwise, the
expectation values do not match. The remaining open parameters $x_i,
\forall i=n+1,\dots d^2$ must be chosen such that
$\rho({\bf{x}})=\rho_{\text{fix}} + \rho_{\text{open}}({\bf{x}}) \geq
0$ forms a positive semidefinite operator and we have defined the
fixed part $\rho_{\text{fix}}=\sum t_i S_i$ and open part
$\rho_{\text{open}}({\bf{x}})=\sum x_i S_i$ of the density
operator. This can be cast into  
the form of a semidefinite program \cite{vandenberghe}, given by
$\min_{(t,{\bf{x}})} t$ subjected to
$F({\bf{x}},t)=\rho({\bf{x}})+t\mathbbm{1} \geq 0$, with solution
$(t^*, {\bf{x^*}})$. If $t^* > 0$, then there exists no parameters
${\bf{x}}$ such that $\rho({\bf{x}})\geq 0$, since otherwise $t^*$ is
not optimal. On the other hand, if $t^* \leq 0$ then $\rho({\bf{x^*}})
\geq 0$, and the expectation value vector ${\bf{t}}$ is quantum mechanical. 

Every semidefinite program has an associated dual program,
cf. Ref.~\cite{vandenberghe}, which reads $\max_Z
  (-\trace(Z \rho_{\text{fix}}))$ subjected to $Z=\sum_{i=1}^n z_i
S_i \geq 0$ and $\trace(Z)=1$. Using the orthogonality of the
operators $S_i$, the objective value can be 
written as $\trace(Z \rho_{\text{fix}})=z^T {\bf{t}}$, and its optimal value is
denoted by $d^*$. Note that both semidefinite programs are 
strictly feasible: If one selects $t>| \min
\lambda(\rho_{\text{fix}})|$, where $\lambda(\cdot)$ denotes the
corresponding eigenvalues, one obtains a strictly positive solution
$F({\bf{x=0}},t)> 0$ for the first program; the operator
$Z=\mathbbm{1}/d > 0$ provides a strictly positive solution for the
second, dual program.  
These conditions ensure strong duality, which states equality of both
programs  $t^*=d^*$, and complementary slackness, which guarantees
that there exist actual parameter sets for
$({\bf{x}}_{\text{opt}},t_{\text{opt}})$, and $Z_{\text{opt}}$ that
attain the solutions $t^*$ and $d^*$ respectively,
cf. Ref. \cite{vandenberghe}. Therefore if ${\bf{t}}$ is not quantum
mechanical it holds that  
\begin{equation}
  -t^*=-d^*=\trace(Z_{\text{opt}}\rho_{\text{fix}})=z^T_{\text{opt}}
  {\bf{t}} < 0,
\end{equation}
Hence every non-quantum mechanical expectation value vector is detected by
the corresponding hyperplane. In contrast, if ${\bf{t}}$ is
quantum mechanical, the weak duality condition \cite{vandenberghe} ensures
that for every feasible solution $z$ of the second program one has
\begin{equation}
  z^T {\bf{t}}=\trace(Z \rho_{\text{fix}})\geq -t^* \geq 0,
\end{equation}
so no quantum mechanical expectation value vector is
detected by the hyperplane. Using the described 1-to-1 correspondence
allows to translate these conditions back to the original operator set
$\{ A_i\}$, which delivers the result.
\end{proof}

\section{Visualization}
\label{section6}

The sets of renormalized expectation values that are determined by the
described inner and outer approximation enclose the set of
quantum mechanical expectation values, and convergence is reached in
the case of infinite spin numbers $j$. In order to visualize each convex
sets, one starts with given expectation values
$u_k$, $k\in\{1,2,3\}$, which fulfill the condition $\sum_k u_k^2 \leq 1$,
which ensures existence of a quantum state, cf. Proposition \ref{prop:
  firstmoment}.  Next, the remaining open parameters $v_k$ must be chosen
according to the conditions from either an approximation set or from the exact 
solution. Using the Casimir identity, given by Eq. \ref{eq: casimir},
allows to further reduce the number of considered parameters $v_k$, and we
choose $v_1$ and $v_2$ in the following. For both approximations, the
search for the extremal values of the parameters $v_k$ can be cast
into the form of a semidefinite program, which can 
be efficiently solved by standard semidefinite program modules
\cite{yalmip}, \cite{sdpt3}; for the exact solutions one needs to
employ a semidefinite program anyway, cf. Sec. \ref{section5}. 
\begin{figure}[ht]
  \centering
  \includegraphics[angle=-90,scale=0.478]{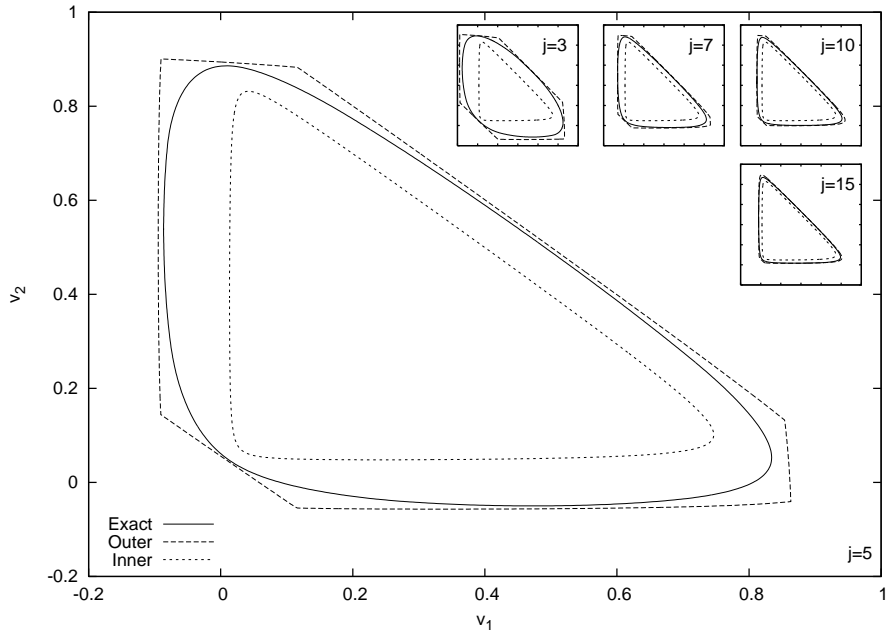}
  \caption{Sets of renormalized expectation values $v_1$ and $v_2$ 
    described by the approximation in comparison with the exact
    solution for the case of $\vec u=(0.1,0.2,0.3)$. The inset figures
    are drawn with the same axis scale, and should demonstrate the
    effect of different total spin numbers $j$ on the different sets.} 
  \label{fig:sets}
\end{figure}
Figure \ref{fig:sets} shows the exact and the approximation sets for
the case of $\vec u=(0.1,0.2,0.3)$ and total spin number $j=5$. The
inset figures visualize the same sets for different spin numbers
$j$, and should demonstrate the corresponding convergence. The outer
approximation seems to describe the actual set of quantum mechanical
expectation values with increasing accuracy as the total spin number
$j$ becomes larger.

\section{Conclusion}
\label{section7}

We have addressed the problem whether a given set of expectation values
can be compatible with expectation values for products of two spin
operators. Those operators, as abstractly introduced as the irreducible
representations of an underlying $\mathfrak{su}(2)$ Lie algebra,
appear for example as the Stokes operators in the quantum optics
literature or as the total angular momentum operators in the case of
atomic ensembles. Because of the particular product structure one can
already impose a strong conditions on the given set of expectation
values, which is summarized in positive semidefiniteness of a
corresponding expectation value matrix that ensures the
Schr\"odinger-Robertson uncertainty principle. Exploiting a particular
representation of the $\mathfrak{su}(2)$ Lie algebra, allows us to
relate the Problem \ref{problem1} to the
Bose-symmetric extension problem for qubits, hence
the following solution to Problem \ref{problem1} is provided: Suppose
that a given set of expectation values   
$M$ satisfies the linear equality imposed by the Casimir identity, one
reconstructs a particular two qubit operator $\rho_j(M)$. This
operator $\rho_j(M)$ must represent a valid two 
qubit density operator of a bipartite qubit system with at least
$2j-2$ Bose-symmetric extensions, otherwise the given
expectation values $M$ disagree with the predictions of quantum
mechanics. Since an exact characterization of two qubit states with a definite
number of extensions is cumbersome, we consider in particular
the two extreme cases. In particular, for large spin numbers we have
presented two different approximating sets. Whenever one finds a
separable two qubit state $\rho_j(M)$, then the expectation values can
be assured to be quantum mechanical. Contrary, if one finds
non-positivity of a particular expectation value matrix
$\tau_j(\rho_j(M)) \not \geq 0$, that depends on the total spin number
$j$ and the reconstructed two-qubit state $\rho_j(M)$, then the
corresponding expectation values are incompatible with the spin
operators. In combination, both tools allow an \emph{approximate operational}
description of the quantum mechanical expectation values. In
particular, the presented method gets better the larger the total spin
number $j$ becomes, and convergence is assured in the limit of infinite
numbers of extensions. In order to provide a feasible
solution for low spin numbers, we characterize the sets of physical
expectation values, similar as in Problem \ref{problem1}, via
hyperplanes. The search for the optimal hyperplane can be cast into the form
of a semidefinite program which can be solved efficiently. 

It remains open whether one can find an operational characterization
of Bose-symmetric extendible two qubit states. In
addition, it might be interesting to further investigate whether
similar ideas can be used if one considers different Lie groups; we
leave this open for future discussions.

\section{Acknowledgments}

We have benefitted from enlightening discussions with many
colleagues, in particular Hauke H\"aseler, Volkher Scholz, Otfried
G\"uhne and Geir-Ove Myhr. This work was funded by the European Union
through the IST Integrated Project SECOQC, the IST-FET Integrated
Project QAP, the NSERC Innovation Platform Quantum Works and the NSERC
Discovery grant. We would like to thank the 
Institute for Quantum Computing and the Institute for Scientific
Interchange Foundation for hospitality and travel support.

\bibliographystyle{unsrt}

\begin{thebibliography}{10}

\bibitem{bell_inequality}
J.~S. Bell.
\newblock On the {E}instein-{P}odolsky-{R}osen paradox.
\newblock {\em Physics}, 1:195, 1964.

\bibitem{nonlocal_boxes}
S.~Popescu and D.~Rohrlich.
\newblock Quantum nonlocality as an axion.
\newblock {\em Foundations of Physics}, 24:379, 1994.

\bibitem{moyal}
J.~E. Moyal.
\newblock Quantum mechanics as a statistical theory.
\newblock {\em Proc. Cambridge Philos. Soc.}, 45:99, 1949.

\bibitem{narcowich}
F.~J. Narcowich.
\newblock The problem of moments in the phase space formulation of quantum
  mechanics.
\newblock {\em J. Math. Phys.}, 28:2873, 1987.

\bibitem{holevo}
A.~S. Holveo.
\newblock {\em Probabilistic and statistical aspects of quantum theory}.
\newblock Amsterdam: North-Holland Pub. Co., 1982.

\bibitem{covariance_matrix}
O.~G\"uhne.
\newblock Characterizing entanglement via uncertainty relations.
\newblock {\em Phys. Rev. Lett.}, 92:117903, 2004.

\bibitem{collins}
Francis~J. Narcowich and R.~F. O\char39{}Connell.
\newblock Necessary and sufficient conditions for a phase-space function to be
  a wigner distribution.
\newblock {\em Phys. Rev. A}, 34:1, 1986.

\bibitem{marcincewicz}
A.~K. Rajagopal and E.~C.~G. Sudarshan.
\newblock Some generalizations of the marcinkiewicz theorem and its
  implications to certain approximation schemes in many-particle physics.
\newblock {\em Phys. Rev. A}, 10:1852, 1974.

\bibitem{korbicz}
J.~K. Korbicz, O.~G\"uhne, M.~Lewenstein, H.~Haeffner, C.~F. Ross, and
  R.~Blatt.
\newblock Generalized spin squeezing inequalities in n qubit systems.
\newblock {\em Phys. Rev. A}, 74:052319, 2006.

\bibitem{korolkova}
N.~Korolkova, G.~Leuchs, R.~Loudon, T.~C. Ralph, and C.~Silberhorn.
\newblock Polarization squeezing and continuous-variable polarization
  entanglement.
\newblock {\em Phys. Rev. A}, 65:052306, 2002.

\bibitem{lorenz}
S.~Lorenz, J.~Rigas, M.~Heid, U.~L. Andersen, N.~L\"utkenhaus, and G.~Leuchs.
\newblock Witnessing effective entanglement in a continuous variable
  prepare-and-measure setup and application to a quantum key distribution
  scheme using postselection.
\newblock {\em Phys. Rev. A}, 74:042326, 2006.

\bibitem{vogel}
E.~Shchukin and W.~Vogel.
\newblock Inseparability criteria for continuous bipartite quantum states.
\newblock {\em Phys. Rev. Lett.}, 95:230502, 2005.

\bibitem{miranowicz}
A.~Miranowicz, M.~Piani, P.~Horodecki, and R.~Horodecki.
\newblock Inseparability criteria based on matrices of moments.
\newblock quant-ph/0605146.

\bibitem{korbicz1}
J.~K. Korbicz and M.~Lewenstein.
\newblock Group-theoretical approach to entanglement.
\newblock {\em Phys. Rev. A}, 74:022318, 2006.

\bibitem{rigas}
J.~Rigas, O.~G\"uhne, and N.~L\"utkenhaus.
\newblock Entanglement verification for quantum-key-distribution systems with
  an underlying bipartite qubit-mode structure.
\newblock {\em Phys. Rev. A}, 73:012341, 2006.

\bibitem{haeseler}
H.~H\"aseler, T.~Moroder, and N.~L\"utkenhaus.
\newblock Testing quantum devices: Practical entanglement verification in
  bipartite optical systems.
\newblock quant-ph/0711.2709.

\bibitem{toth}
G.~T\'oth, C.~Knapp, O.~G\"uhne, and H.~J. Briegel.
\newblock Optimal spin squeezing inequalities detect bound entanglement.
\newblock quant-ph/0702219.

\bibitem{hall}
B.~C. Hall.
\newblock {\em Lie groups, Lie algebras, and representations: an elementary
  introduction}.
\newblock Springer-Verlage New York, 2003.

\bibitem{wang}
X.~Wang and K.~M$\o$lmer.
\newblock Pairwise entanglement in symmetric multi-qubit systems.
\newblock {\em Euro. Phys. J. D}, 18:385, 2002.

\bibitem{doherty}
A.~C. Doherty, P.~A. Parrilo, and F.~M. Spedalieri.
\newblock Complete family of separability criteria.
\newblock {\em Phys. Rev. A}, 69:022308, 2004.

\bibitem{doherty1}
A.~C. Doherty, P.~A. Parrilo, and F.~M. Spedalieri.
\newblock Distinguishing separable and entangled states.
\newblock {\em Phys. Rev. Lett.}, 88:187904, 2002.

\bibitem{terhal}
B.~M. Terhal, A.~C. Doherty, and D.~Schwab.
\newblock Symmetric extensions of quantum states and local hidden variable
  theories.
\newblock {\em Phys. Rev. Lett.}, 90:157903, 2003.

\bibitem{morodersym}
T.~Moroder, M.~Curty, and N.~L\"utkenhaus.
\newblock One-way quantum key distribution: Simple upper bound on the secret
  key rate.
\newblock {\em Phys. Rev. A}, 74:10, 2006.

\bibitem{wolf}
M.~M. Wolf and D.~P\'erez-Garci\'a.
\newblock Quantum capacities of channels with small environment.
\newblock {\em Phys. Rev. A}, 75:4, 2007.

\bibitem{peres}
A.~Peres.
\newblock Separability criterion for density matrices.
\newblock {\em Phys. Rev. Lett.}, 77:1413, 1996.

\bibitem{entanglement_witness}
M.~Horodecki, P.~Horodecki, and R.~Horodecki.
\newblock Separability of mixed states: necessary and sufficient conditions.
\newblock {\em Phys. Lett. A}, 223:1, 1996.

\bibitem{wangsanders}
X.~Wang and B.~C. Sanders.
\newblock Spin squeezing and pairwise entanglement for symmetric multiqubit
  states.
\newblock {\em Phys. Rev. A}, 68:012101, 2003.

\bibitem{stokten}
J.~K. Stockton, J.~M. Geremia, A.~C. Doherty, and H.~Mabuchi.
\newblock Characterizing the entanglement of symmetric many-particle spin-1 / 2
  systems.
\newblock {\em Phys. Rev. A}, 67:022112, 2003.

\bibitem{korbicz2}
J.~K. Korbicz, J.~I. Cirac, and M.~Lewenstein.
\newblock Spin squeezing inequalities and entanglement of n qubit states.
\newblock {\em Phys. Rev. Lett.}, 95:120502, 2005.

\bibitem{vandenberghe}
L.~Vandenberghe and S.~Boyd.
\newblock Semidefinite programming.
\newblock {\em SIAM Review}, 38:49, 1996.

\bibitem{yalmip}
J.~L{\"o}fberg.
\newblock {YALMIP} : A toolbox for modeling and optimization in {MATLAB}.
\newblock In {\em Proceedings of the {CACSD} Conference}, pages 284--289,
  Taipei, Taiwan, 2004.
\newblock Available from
  \begin{tt}http://control.ee.ethz.ch/\~{}joloef/yalmip.php\end{tt}.

\bibitem{sdpt3}
K.~C. Toh, R.~H. Tutuncu, and M.~J. Todd.
\newblock Sdpt3--a matlab software package for semidefinite programming.
\newblock {\em Optimization Methods and Software}, 11:545--581, 1999.
\newblock Available from
  \begin{tt}http://www.math.nus.edu.sg/~mattohkc/sdpt3.html\end{tt}.

\end{thebibliography}

\end{document}